\documentclass[prb,preprint,showpacs]{revtex4}

\usepackage{graphicx}
\usepackage{amsmath}

\begin{document}
Journal Reference: Physical Review B {\bf 82}, 245205 (2010)
\title{Strong electron correlations in FeSb$_2$: An optical investigation  and comparison with RuSb$_2$}
\author{A. Herzog}
\author{M. Marutzky}
\author{J. Sichelschmidt}\email{Sichelschmidt@cpfs.mpg.de}
\author{F. Steglich}
\affiliation{Max-Planck-Institut f\"ur Chemische Physik fester Stoffe, N\"othnitzer Str. 40, D-01187
Dresden, Germany}
\author{S. Kimura}
\affiliation{UVSOR Facility, Institute for Molecular Science, Okazaki 444-8585, Japan}
\author{S. Johnsen}
\author{B.B. Iversen}
\affiliation{Department of Chemistry, University of Aarhus, Langelandsgade 140, 8000 Aarhus C, Denmark}

\begin{abstract}
We report investigations of the optical properties of the narrow gap semiconductor FeSb$_2$ in comparison with the structural homolog RuSb$_2$. In the infrared region the latter shows insulating behavior in whole investigated temperature range ($10-300{\,\rm K}$) whereas the optical reflectivity of FeSb$_2$ shows typical semiconductor behavior upon decreasing the temperature. The conduction electron contribution to the reflectivity is suppressed and the opening of a direct and an indirect charge excitation gap in the far-infrared energy region is observed. Those gap openings are characterized by a redistribution of spectral weight of the optical conductivity in an energy region much larger than the gap energies indicating that strong electron-electron correlations are involved in the formation of the charge gap. Calculations of the optical conductivity from the band structure also provided evidence for the presence of strong electronic correlations. Analyzing the spectra with a fundamental absorption across the gap of parabolic bands yields a direct gap at 130~meV and two indirect gaps at 6 and 31 meV. The strong reduction in the free-carrier concentration at low energies and low temperatures is also reflected in a change in the asymmetry of the phonon absorption which indicates a change in the phonon-conduction-electron interaction. 
\end{abstract}

\pacs{71.28.+d, 71.27.+a, 78.20.--e}

\maketitle

\section{Introduction}
Strong electron correlations may lead to semiconducting behavior with a narrow charge excitation gap. This may be realized when a broad conduction band weakly hybridizes with a flat band arising from local atomic orbitals and, involving the Kondo exchange interaction between conduction electrons and local electrons, the picture of a "correlated semiconductor" or "Kondo insulator" may be applied.\cite{aeppli92a} Such hybridization gap formation is reported for a small group of rare-earth containing intermetallic compounds where the Kondo-like interaction between conduction electrons and the well localized $f$-electrons leads to a insulating ground state. This process essentially accounts for the unconventional temperature dependencies for the gap formation which is reflected in a number of temperature dependent, narrow gap semiconductor properties.\cite{fisk95a,riseborough00a} In contrast, a common semiconductor picture characterizes the gap with a simple thermal charge carrier activation across rigid bands.

Infrared optical spectroscopy can provide important information about the gap formation process. In case of a Kondo insulator the gap formation may occur at much lower temperatures than the temperature corresponding to the gap energy. This behavior is not expected for gaps between rigid bands and has been reported, for example, for the prototype Kondo insulator Ce$_3$Bi$_4$Pt$_3$.\cite{bucher94a} Additionally, on cooling, the optical spectral weight is shifted from below the gap to energies an order of magnitude larger, pointing to strong electronic correlations involved in the charge gap formation.\cite{schlesinger93a, rozenberg96a}
Recent calculations within the periodic Anderson model show that the optical behavior of Kondo insulators is governed by an \textit{indirect} energy gap in the mid-infrared region which is associated with the gap in the density of states (DOS) and a \textit{direct} gap at higher energies that controls the anomalous redistribution of spectral weight.\cite{franco09a}

The narrow band gap semiconductor FeSb$_2$ (orthorhombic marcasite-type structure \cite{hulliger63a, fan72a}) was discussed as a possible non-rare-earth containing correlated semiconductor,\cite{petrovic05a} besides the extensively discussed FeSi.\cite{schlesinger93a, mandrus95a} Similar to FeSi, a correlated semiconductor scenario for FeSb$_2$ would be favored by the fact that it gets semiconducting and diamagnetic in the same temperature region. Also, the charge excitation spectra of either these compounds display features of a hybridization gap and strong electron correlations.\cite{perucchi06a, schlesinger93a} However, in the case of FeSi, it still seems not settled whether strong electron correlations could be inferred from an anomalous spectral weight redistribution in the optical conductivity. Such conjecture was drawn from reflectivity measurements that may not allow to provide precise enough data for a proper analysis of the spectral weight redistribution, see the discussion in Ref. \onlinecite{menzel09a}. Moreover, high-resolution ARPES data of FeSi revealed narrow bands which could be described by simple band structure calculations and without invoking Kondo insulator physics.\cite{klein08a} An alternative picture of a nearly ferromagnetic semiconductor successfully reconciles the experimental results with local-density approximation band structure calculations \cite{lukoyanov06a} as well as with treating temperature dependent correlations within a dynamical mean-field approximation.\cite{kunes08a}
 
Perucchi et al. reported first optical investigations of FeSb$_2$ using samples which remain metallic along the $b$-axis at low temperatures.\cite{perucchi06a} They observed an unconventional gap formation below 30~meV and $T=80$~K. The spectral weight is not only redistributed toward energies higher than 1~eV but also piles up as a continuum of excitations in the far-infrared spectral region. Therefore, from these optical measurements the application of a correlated semiconductor description for FeSb$_2$ remained an open issue, and the presence of in-gap localized states in connection with lattice dynamical effects were suggested to explain the unusual spectral weight redistribution. 

More clear evidence for strong electron-electron correlations in FeSb$_2$ comes from samples which are characterized by a semiconducting transport behavior along \textit{all} crystalline axes and where, at around 10~K, a colossal Seebeck coefficient $S$ is observed.\cite{bentien07a} The latter indicates high density of states at the edges of a gap which could be formed by a weak hybridization of a broad conduction band with a narrow $d$-band. Concomitantly to the colossal $S$ a thermally activated type of charge transport with a small energy gap of 6~meV is observed in the same temperature region. Isostructural RuSb$_2$, considered as a conventional band semiconductor with a two order of magnitude smaller thermopower, has shown to be a reference material without contributions from narrow bands and correlated electrons.\cite{sun09a} 

We investigated the charge dynamics of FeSb$_2$ using single crystalline samples which show a colossal Seebeck coefficient along with semiconducting transport properties for all crystal axes.\cite{bentien07a}
We compare the observed temperature dependent low-energy suppression of the optical response with the corresponding optical response of the structural homologue RuSb$_2$ which shows a more conventional behavior such as, for instance, a constant diamagnetism in the whole investigated temperature range.\cite{sun09c} 
%
%
%
%
\section{Experiment and Results}
FeSb$_2$ and RuSb$_2$ single crystals with large (110) facets ($\sim3\times3$~mm$^2$) were grown using a self-flux method as described in Refs. \onlinecite{bentien07a} and \onlinecite{sun09a}. After well polishing the sample surfaces (0.3 mm grain) we performed near-normal incidence measurements of the optical reflectivity $R(\omega)$ with electric field $\mathbf{E}$ polarized parallel to the crystallographic $\mathbf{c}$-axis.
A rapid-scan Fourier spectrometer of Michelson and Martin-Puplett type was used for energies between 3\,meV and 3\,eV $(2\,{\rm K}<T<300\,{\rm K})$ and for higher energies at $T=300$\,K only. Synchrotron radiation extended the energy range from 1.2\,eV up to 30\,eV.\cite{fukui01a} For determination of $R(\omega)$ the samples were coated \emph{in-situ} with gold and then used for measuring the reference spectrum. 
%
%
\subsection{Reflectivity}
\begin{figure}
\centering
\includegraphics[width=10cm]{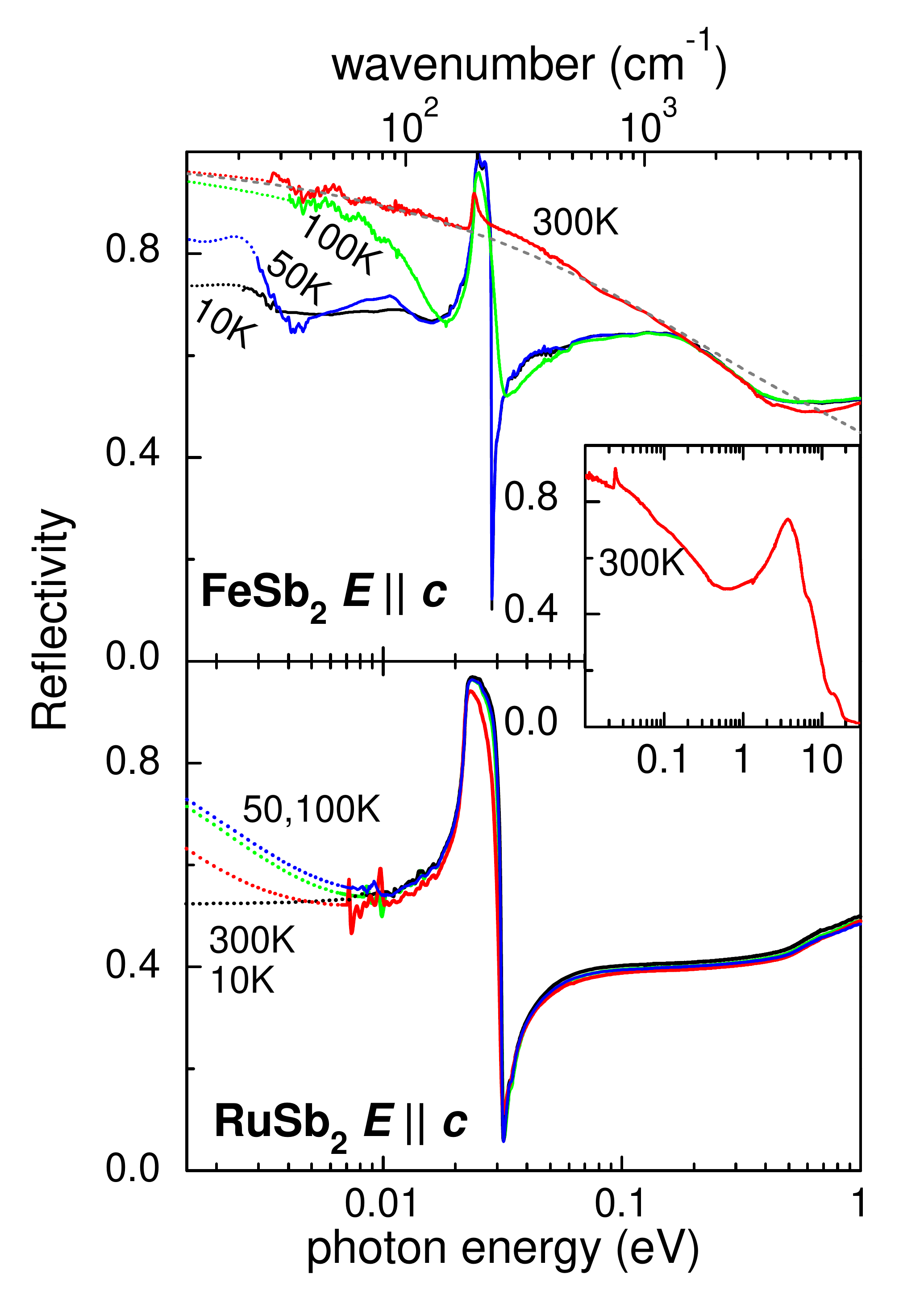}
\caption{Reflectivity $R(\omega)$ of FeSb$_2$ and RuSb$_2$ with $\mathbf{E} \ || \ \mathbf{c}$ in the (110) plane at various temperatures as indicated. Dotted lines below  $\approx3$~meV correspond to extrapolations of the optical conductivity to the zero-energy DC values (see Fig.2). Dashed line shows a Drude model with $\hbar\omega_p=2.5$~eV and $\hbar\gamma=1$~eV. Insert displays $R(\omega)$ for FeSb$_2$ ($\mathbf{E} \ || \ \mathbf{c}$) in the high-energy region.} \label{reflparc}
\end{figure}

The reflectivity $R(\omega)$ of FeSb$_2$ and RuSb$_2$ is depicted at various temperatures in Fig. \ref{reflparc}. The dotted lines in the low-energy part indicate $R(\omega)$ values which correspond to the optical-conductivity extrapolation toward zero energy as discussed below. The high-energy part covering the electronic interband transition peaks is shown in the insert frame for FeSb$_2$ at $T=300$~K. The pronounced narrow structure near 200~cm$^{-1}$ is due to \textit{one} phonon absorption as expected from a group theoretical analysis for this geometry of excitation. It is located on a broad background and appears at almost the same energy for both compounds reflecting their same structural symmetry. For FeSb$_2$, the broad background strongly depends on temperature and energy which agrees with the much larger conductivity of FeSb$_2$ (Ref. \onlinecite{sun09a}) leading to a plasma edge of a free carrier excitation with a characteristic gradual increase in $R(\omega)$ toward low energies. In Fig. \ref{reflparc} the dashed line displays this excitation for the 300~K data with a Drude model, using a free-carrier plasma energy of 2.5~eV, a damping of 1~eV, and $\epsilon_\infty=38$. Assuming a free electron mass these parameters correspond to a density of $4\cdot10^{27}\rm{m^{-3}}$ and a mobility of $1\cdot10^{-3}\rm{m^2/Vs}$ which agree well with values derived from Hall effect and resistivity measurements.\cite{sun09a, sun09b} With decreasing temperature the energy of the Drude edge strongly decreases, reflecting the reduction in the charge-carrier density as seen in the Hall coefficient.\cite{sun09a} The $R(\omega)$ spectra of RuSb$_2$ are almost temperature independent and do not show any free carrier contribution. Compared to FeSb$_2$ this is expected from its more resistive behavior \cite{sun09a} and indicates that the energy scales of charge excitations exceed those of FeSb$_2$.
It is worth to note the similarity of the $R(\omega)$ spectra at $T=10$~K for both compounds. This observation well agrees with the similarity in their Hall coefficient, i.e., their charge-carrier concentration at low temperatures.\cite{sun09a} \\
%
%
\subsection{Optical conductivity}
\begin{figure}
\centering
\includegraphics[width=10cm]{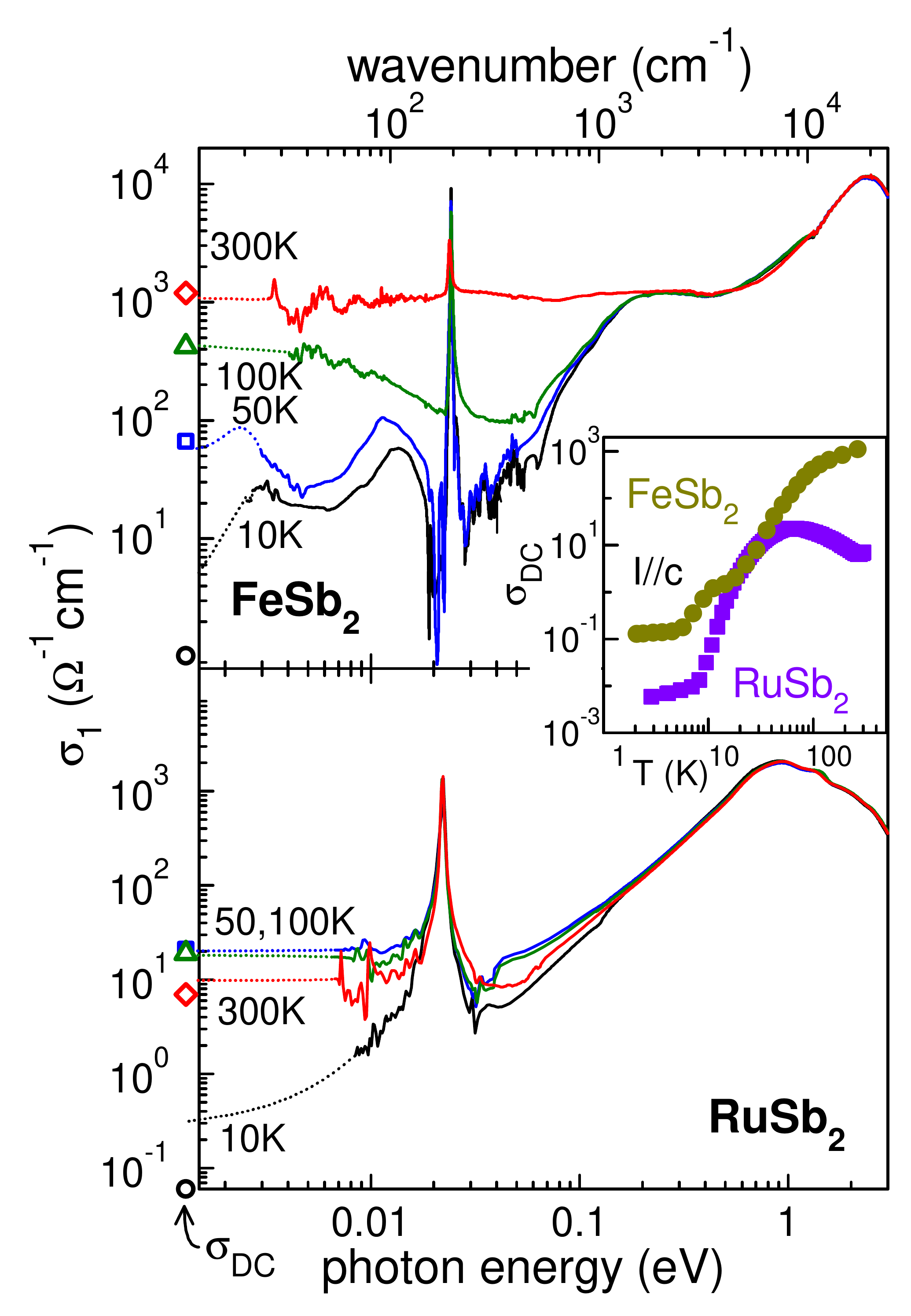}
\caption{Optical conductivity $\sigma_1(\omega)$ of FeSb$_2$  and RuSb$_2$ with $\mathbf{E} \ || \ \mathbf{c}$ in the (110) plane at various temperatures as indicated. Symbols indicate the DC-values $\sigma_{\rm DC}=\sigma_1(\omega=0)$ at the respective temperatures. Insert displays temperature dependence of $\sigma_{\rm DC}$ (same units, DC current along $\mathbf{c}$).} \label{optcondparc}
\end{figure}

The dissipative part $\sigma_1(\omega)$ of the complex optical conductivity is depicted in Fig. \ref{optcondparc}. $\sigma_1(\omega)$ was calculated from the reflectivity by a Kramers-Kronig transformation after appropriate extrapolations. The high-energy region was extrapolated with a free-electron approximation of the form $R(\omega)\propto\omega^{-4}$. Towards zero energy we used a Drude-Lorentz model extrapolating the low-energy $R$ data (dashed lines in Fig. \ref{optcondparc}) towards zero energy. This is done in such a way that the $\sigma_1(\omega)$ data fit to the zero-energy conductivity $\sigma_{\rm DC}$ (symbols at the ordinates of Fig. \ref{optcondparc}) which was measured at the same samples of FeSb$_2$ and RuSb$_2$. Uncertainties of the $\sigma(\omega)$ spectra due to variations in the low-energy extrapolation of $R(\omega)$ are found to be negligible for energies above 6~meV. Below 6~meV, for the data of FeSb$_2$ at 10 and 50~K, it turns out that the $\omega\rightarrow0$ extrapolations  require an excitation at $\hbar\omega\approx2$~meV which is just below the low-energy end of the measured data. Without such a model and if using a only Drude tail for the steep increase in $R(\omega<4\,\rm{meV})$ the optical-conductivity extrapolation would not agree with $\sigma_{\rm DC}$ within one order of magnitude even if to consider the errors due to the difficulties when measuring very low DC-conductivities. A similarly small energy scale also characterizes the electrical resistivity at the lowest temperatures ($<10$~K) and was discussed to be probably due to extrinsic impurity states.\cite{bentien07a} \\       
With these extrapolations of $R(\omega)$ the resulting Kramers-Kronig derived optical conductivity $\sigma_1(\omega)$ is characterized as follows: 
both compounds display a behavior typical for semiconductors, {i.e.} a strong suppression of $\sigma_1(\omega)$ at the band absorption edge. The spectra nicely reflect the semiconducting gaps which are also involved in the DC-conductivity (shown in inserted frame) as thermally activated excitations across band gaps. For FeSb$_2$ $\sigma_{\rm DC}(T)$ indicates a band gap of $\simeq30$~meV (Ref. \onlinecite{bentien07a}) which is found in the optical response as an indirect gap at low temperatures as will be analyzed below. It is filled up by thermal excitations at 300~K, leading to a flat $\sigma_1(\omega)$. Below a photon energy of $\approx10$~meV and at low temperatures $\sigma_1(\omega)$ suffers an additional small suppression. This observation corresponds to an energy scale in $\sigma_{\rm DC}$ which can be described by a small activation energy of 6~meV between $T=5-15$~K and which is related to a colossal Seebeck coefficient as well as a small Hall coefficient.\cite{bentien07a} Also this feature presumably corresponds to an indirect gap, see discussion below.
$\sigma_{\rm DC}(T)$ of RuSb$_2$ is characterized by a metallic behavior in the region $50-280$~K and a thermal activation across energy gaps of 14~meV ($T<50$~K) and 293~meV ($T>280$~K).\cite{sun09a} It is the latter gap that mainly characterizes $\sigma_1(\omega)$ showing at a similar energy a strong decrease which corresponds to an indirect gap and which depends only weakly on temperature up to 300~K. The similarity of the DC conductivities of RuSb$_2$ and FeSb$_2$ around $T=30$~K corresponds to a similar optical conductivity at $T=50$~K and below 0.1~eV.\\
%
%
\subsection{Charge Gap and Interband Transitions}
\begin{figure}
\centering
\includegraphics[width=10cm]{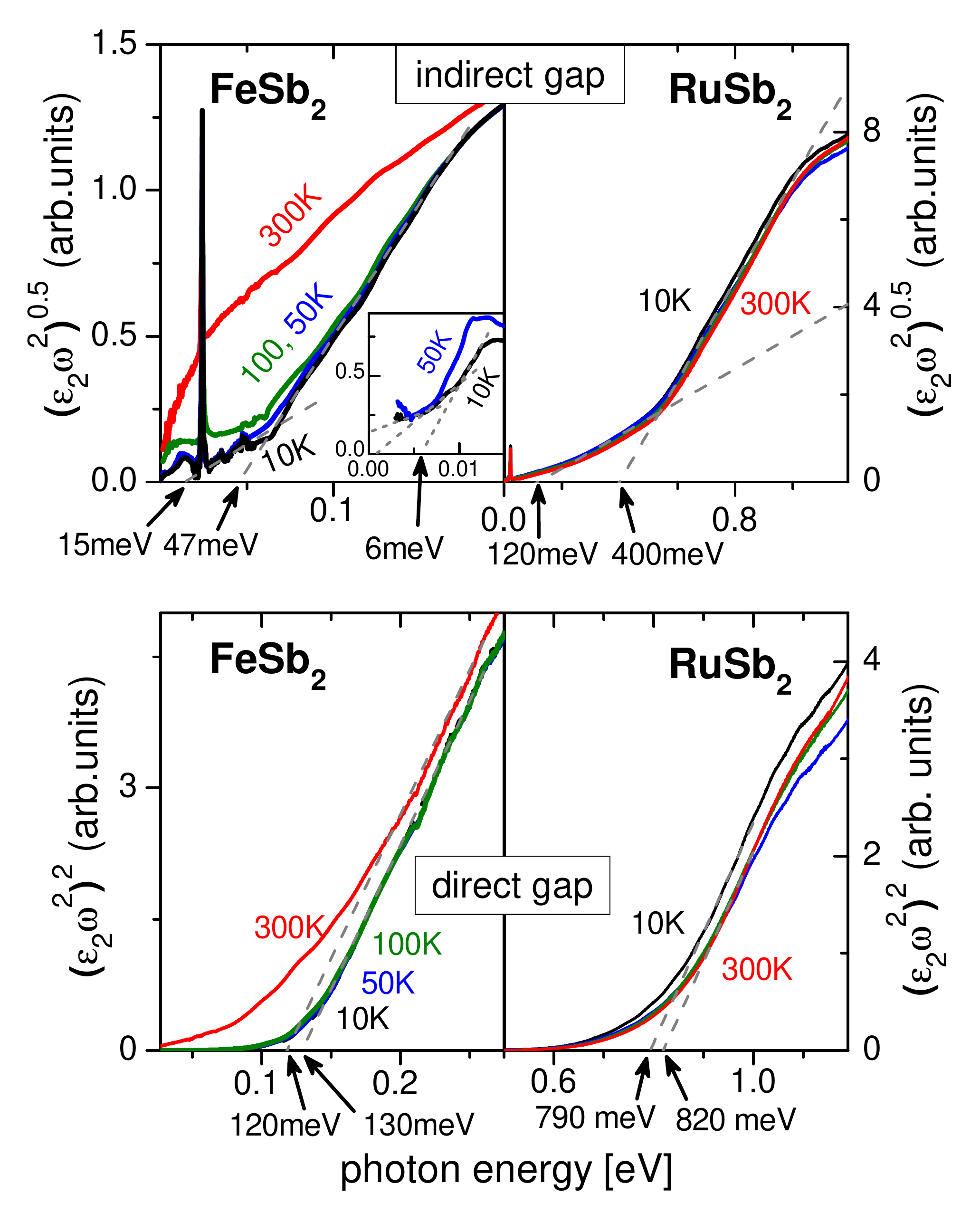}
\caption{Energy dependence of the absorption coefficient $K$ expressed as $(K\omega)^m\propto(\epsilon_2\omega^2)^m$ for $\mathbf{E} \ || \ \mathbf{c}$. According to Eq.(\ref{gap}) the intersection of the linear fits (dashed lines) with the energy axis reveals the indicated values of $E^*$ which are related to indirect gaps ($m=0.5$, top frame) or to direct gaps ($m=2$, bottom frame). The inset in the upper frame depicts the low-energy part where three branches of indirect absorptions are indicated by dashed lines.} 
\label{absorbanz}
\end{figure}

A transition across the band gap $E_g$ of a semiconductor causes an absorption edge in the optical response. Whether such transitions are direct or indirect (involving phonon emission and absorption) can be evaluated by inspecting the behavior of the absorptive of the dielectric function $\epsilon_2(\omega)$ for which in case of parabolic bands and near to the band edge one can derive:
\begin{equation} \label{gap}
(\epsilon_2\omega^2)^m \propto \hbar\omega - E^*,
\end{equation}
with $m = 2$ and $E^*=E_{\rm{g}}^{\rm{dir}}$ for a direct transition and $m = 1/2$ and $E^*=E_{\rm{g}}^{\rm{ind}}\pm E_{\rm{phonon}}$ for an indirect transition.\cite{dressel02a, menzel09a, yu05a} The gap values may then directly be obtained by plotting $(\epsilon_2\omega^2)^m$ of Eq.(\ref{gap}) versus the photon energy as shown in Fig. \ref{absorbanz}.
In this way we determined at $T=10$~K a phonon assisted indirect gap of $E^*\approx(31\pm16)$meV for FeSb$_2$ and $E^*\approx(260\pm140)$meV for RuSb$_2$, which are both in good agreement with transport-measurement results.\cite{sun09a} As shown in the inset 
of Fig. \ref{absorbanz} the low-temperature suppression of $\sigma_1(\omega)$ below $\approx10$~meV corresponds to indirect gap features in $(\epsilon_2\omega^2)^{0.5}$: three upper branches $E^*=E_{\rm{g}}^{\rm{ind}}+E_{\rm{phonon}}$ are shown by dashed lines. Interestingly, the most upper branch corresponds to an energy of $\approx6$~meV which agrees with the energy attributed to the colossal Seebeck coefficient.\cite{bentien07a} Very similar observations are reported for FeSi where linear slopes in $(\epsilon_2\omega^2)^{0.5}$ are related to indirect transitions with assisting emissions of acoustic phonons.\cite{menzel09a}
The direct gap values of $E^*\approx130$meV for FeSb$_2$ and $E^*\approx790$meV for RuSb$_2$ show a small temperature dependence which is probably due to thermal-expansion effects \cite{varshni67a} or experimental uncertainties.\\
Besides the gap features the optical response shows broad interband electronic transitions. A Lorentz model was fitted to $\sigma_{1}(\omega)$ with a Lorentz oscillator for each transition $j$, 
\begin{equation} \label{lorentz}
\sigma^j(\omega) = \omega^2_{pj}\epsilon_{0}\frac{\omega}{i({\omega_{0j}}^2-\omega^2)+\omega\tau_j} \text{.}
\end{equation}
with results displayed in Table \ref{fitpara10K}. Here, $\hbar\omega_{0j}$ is the transition energy of the oscillator $j$, $\hbar\tau^{-1}_j$ the scattering rate or damping constant, and $\hbar\omega_{p}={Ne^2}/ {\varepsilon_0 m^*}$ the effective plasma energy, where $N$ is the concentration of the conduction electrons and $m^*$ their effective mass.

\begin{table}[h!]
\includegraphics[width=8cm]{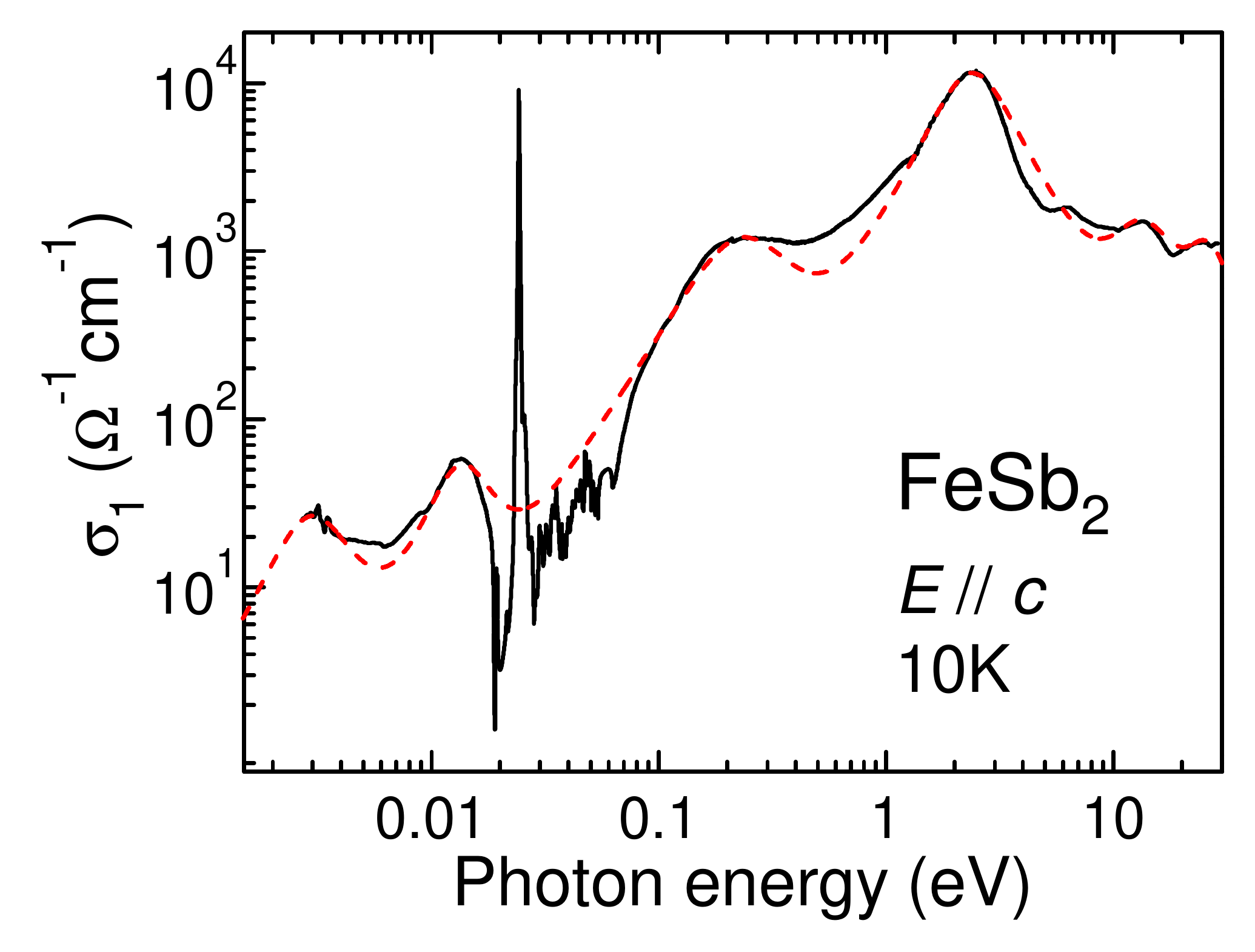} \hfill
\begin{tabular}{|c|c|c|c|} \hline
   \ \ $j$ \ \ & \ \  $\hbar\omega_{0j}$ [eV]\ \ & \ \ $\hbar\omega^*_{pj}$ [eV] \ \  &  \ \ $\hbar\tau^{-1}_j$ [eV] \ \  \\ \hline
   1 & 0.003 & 0.022 & 0.0026 \\
  2 & 0.014 & 0.057 &  0.0093 \\
  3 & 0.23 & 1.5 & 0.27 \\ 
  4 & 2.4 & 13.1 & 2 \\
  5 & 13.4 & 9.2 & 9.8 \\ 
  6 & 25.9 & 9.2 & 13.7 \\ \hline
\end{tabular}
\caption{The figure shows a Lorentz fit (dashed line, Eq. \ref{lorentz}) to $\sigma_{1}$ ($\omega$, 10 K, $\mathbf{E} \ || \ \mathbf{c}$) with parameters listed in the table.} \label{fitpara10K}
\end{table}

Calculations applying the LDA+U method\cite{lukoyanov06a} predict a small indirect gap as well as a larger gap at 300 meV between $3d(3z^2-r^2)$ states in the valence band and $3d(x^2-y^2)$ states in the conduction band. This gap value is consistent with the excitation $j=3$ listed Table \ref{fitpara10K} and probably takes place from the occupied 4$p$ states of Sb to the unoccupied $3d(x^2-y^2)$ states of Fe since a $d-d$ transition is optically forbidden. 

\begin{figure}[h!]
\centering
\includegraphics[width=10cm]{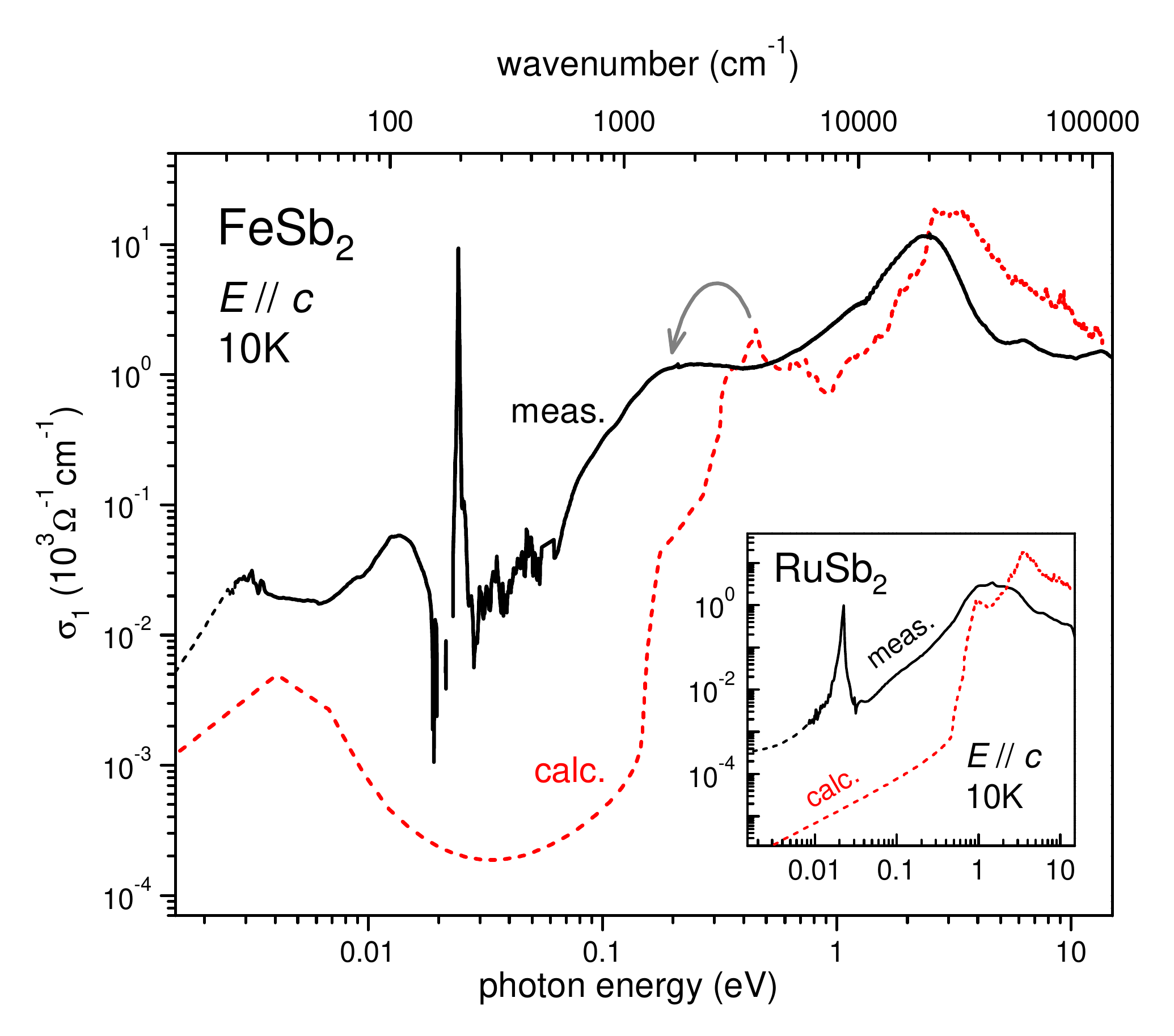}
\caption{Calculation of the optical conductivity (dashed lines) performed for FeSb$_2$ and RuSb$_{2}$ (inset). Arrow indicates a considerable shift between calculation and experiment.} \label{rechnungen}
\end{figure}

For a more direct comparison we calculated the optical conductivity from the band structure of FeSb$_2$ and RuSb$_{2}$. The calculation was performed using a full potential linearized augmented plane wave plus the local orbital including spin-orbit coupling implemented in the WIEN2K code.\cite{blaha90a, blaha01a} The result is depicted in Fig. \ref{rechnungen} and shows for both compounds a reasonable agreement with the optical conductivity data. For FeSb$_{2}$, in the low energy region the experimental peaks ($j=1$ and $j=2$, Tab. \ref{fitpara10K}) which correspond to indirect gap transitions (see Fig. \ref{absorbanz}) have no direct counterpart in the calculation. Instead, probably because of the poor resolution ($>10$~meV) of the calculation, a single calculated low energy peak structure appears and  corresponds to a small density of (''in-gap``) states with mostly Sb-$p$ character at the top of the valence band (R-point of the band structure).\cite{lukoyanov06a}   

In RuSb$_{2}$ low energy peaks are not present in the experimental data in agreement with the calculation, see inset of Fig. \ref{rechnungen}. Also, a considerable absorption is correctly reproduced for energies above $\approx1$~eV. In contrast to RuSb$_{2}$, a strong absorption in FeSb$_{2}$ occurs already at lower energies and leads to the above mentioned peak at 0.23~eV ($j=3$). As indicated by the bended arrow a considerable shift of this peak from the peak in the band calculation is observed. This points to a strong renormalization factor due to the electronic correlations of the Fe 3$d$ state. Such conclusion was drawn for Ytterbium compounds from the consistency between calculated renormalized peaks and experimental peaks.\cite{kimura09b} For FeSb$_2$ a mass enhancement of about 14 free-electron masses was inferred from the electronic specific heat and a one-band carrier model \cite{sun09a} which may point out the relevance of strong correlations. 
However, it was shown for FeSi that the band structure close to the Fermi level observed by ARPES can be described by a simple self-energy correction to a DFT band-structure calculation.\cite{klein08a} This is at variance with previous conjectures from band-structure calculations in FeSi.\cite{mattheiss93a} Hence, at least for FeSi it seems that strong local interactions may not be required to explain experimental results as, for example, a low-temperature mass-enhancement of about 30 free-electron mass seen in specific-heat measurements.\cite{chernikov97a}

The calculated peaks at around 3~eV are consistent with the data of FeSb$_2$ ($j=4$, Table \ref{fitpara10K}) and RuSb$_{2}$. They could be due to transitions from 3$d$ states with large DOS $\approx 1$~eV below the Fermi energy to 4$p$ states with relatively large partial DOS at 2~eV above the Fermi energy. 

%
%
\subsection{Spectral weight}
Whether or not strong electron correlations are relevant for the optical gap formation may be indicated by considering the transfer of spectral weight to energies above the gap.\cite{schlesinger93a,rozenberg96a}
\begin{figure}[h!]
\centering
\includegraphics[width=8cm]{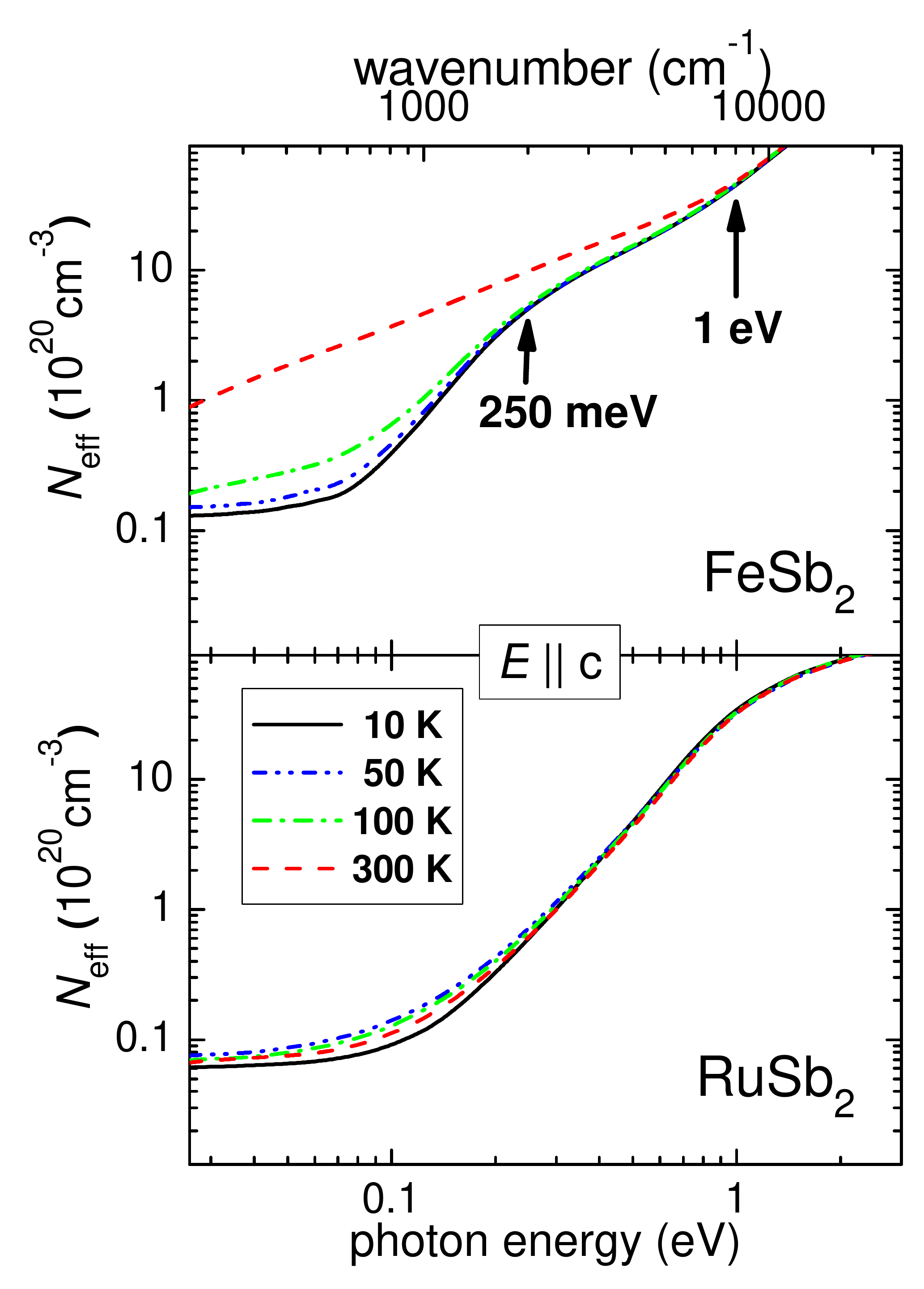}
\caption{The effective number of electrons $N_{eff}(\omega)$ contributing to the absorption for $\mathbf{E} \ || \ \mathbf{c}$ up to energies of $\omega$. Phonon contributions were subtracted from the spectral weight.} \label{spektralgewicht}
\end{figure}
As discussed above, in contrast to RuSb$_{2}$, for FeSb$_2$ strong electron correlations could be relevant for the optical properties. A comparison of both compounds regarding the temperature dependence of the optical spectral weight $N_{eff}(\omega)$ is shown in Fig. \ref{spektralgewicht}. $N_{eff}(\omega)$ is a measure of the effective number of electrons $N_{eff}$ which are contributing to the absorption up to a particular energy $\omega$ and was calculated according the $f$ sum rule for the optical conductivity \cite{dressel02a}
\begin{equation} \label{sumrule}
\int^{\infty}_{0}\sigma_1(\omega)d\omega = \frac{{\omega_p}^2}{8}=\frac{\pi N e^2}{2m^{*}}
\end{equation}
by integrating $\sigma_{1}$ up to $\omega$.  

Clearly, the behavior of $N_{eff}(\omega)$ of FeSb$_2$ and RuSb$_{2}$ is different in respect to the charge-gap formation at low temperatures. 
Whereas  for RuSb$_{2}$ (direct gap energy $\approx 0.8$~eV, Fig. \ref{absorbanz}) the recovery of the spectral weight loss is seen at $\approx 0.8$~eV the recovery for FeSb$_{2}$ (direct gap energy $\approx 0.13$~eV, Fig. \ref{absorbanz}) occurs at $\approx 1$~eV, i.e., the spectral weight upon gap-formation is distributed over energies much larger than the gap-energy. This behavior is consistent with the expectations when strong electronic correlations are involved in the charge-gap formation. \cite{schlesinger93a, rozenberg96a}      
In addition, at an energy of 0.25~eV the spectral-weight transfer of a gap opening for temperatures below 100~K seems to be recovered.
If this feature could be related to the above-mentioned low-energy indirect excitations (6 and 31~meV, Fig. \ref{absorbanz}) another fingerprint of strong electron correlations may be identified. In this respect one should note that the strongly enhanced thermopower in FeSb$_2$ at temperatures below 30~K is related to strongly renormalized effective masses of bands at the Fermi level and, therefore, ascribed to the presence of electron-electron correlations.\cite{bentien07a, sun09b}
  
%
\subsection{Phonons}

As reported in Ref.\cite{racu07a} the Raman and infrared active phonons of FeSb$_{2}$ single crystals were investigated in detail by a factor group analysis yielding a full mode assignment. Here, we consider the effect of the electron-phonon interaction to the lineshape of the phonon mode $B_{\rm{1u}}$ in more detail by analyzing the lineshape with an extension of the Lorentz model, the Fano model. It treats the absorption of a discrete state coupled to a continuum, which is, for example, realized by a phonon-electron coupling \cite{fano61a, damascelli97a}:
\begin{equation} \label{fano}
{\sigma^j}_{\rm Fano}(\omega) = i\frac{\epsilon_0\omega_{pj}^{*}}{{\tau_j}^{-1}q^2}\frac{(q_j + i)^2}{x_j(\omega)+i}
\end{equation}
with
\begin{equation} \label{fano2}
x_j(\omega) = \frac{\omega_{0j}^2 - \omega^2}{\tau_j^{-1}\omega} \text{.}
\end{equation}
While in the Lorentz model the lineshape of the dielectric function $\varepsilon_2 = \sigma_1/\varepsilon_0\omega$ appears symmetric, the Fano parameter $q$ causes an asymmetry of the lineshape. For $|q|\rightarrow \infty$ the Fano model passes over into the Lorentz model.
 
%
\begin{figure}
\centering
\includegraphics[width=10cm]{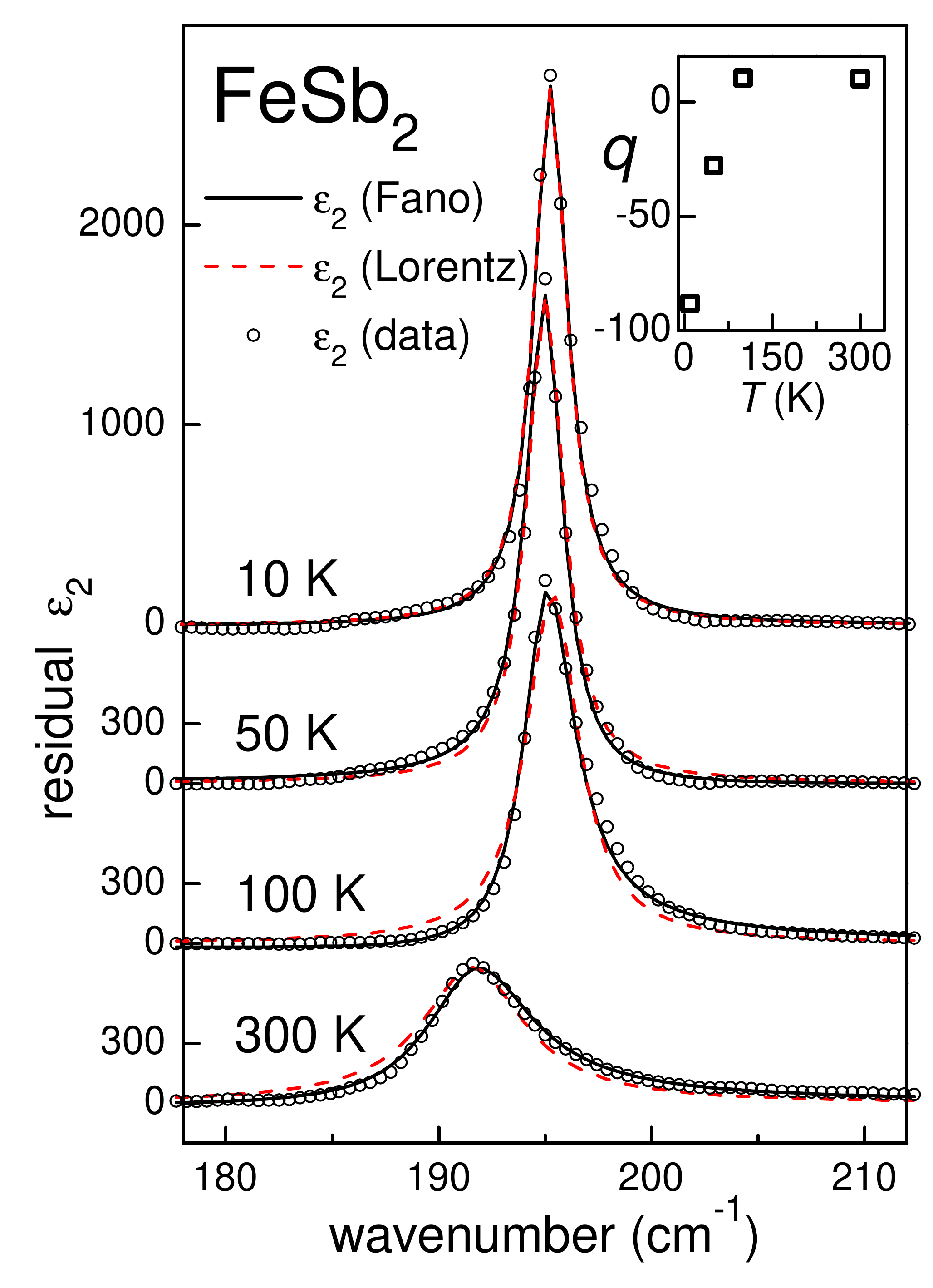}
\caption{ Temperature variation in the residual dielectric function for $\mathbf{E} \ || \ \mathbf{c}$ (circles) which was obtained  after subtraction of purely electronic contributions (Table \ref{fitpara10K}). The displayed phonon absorption $B_{\rm{1u}}$ was fitted with a Lorentz- (dashed lines, Eq. \ref{lorentz}) and a Fano model (solid lines, Eq. \ref{fano}) with the asymmetry parameter $q$ as displayed in the inset.}\label{phonon}
\end{figure}
After subtracting the electronic contribution we obtained the ``residual $\varepsilon_{2}$'' shown in Fig. \ref{phonon} in an energy range where the $B_{\rm{1u}}$ phonon mode is nicely visible. Between 10 and 300~K the lineshape of the phonon changes from symmetric to asymmetric which means that at high temperatures the lineshape fits by a Lorentzian are poor whereas the Fano model is more suitable. 
As shown in the inset of Fig. \ref{phonon} the absolute value of the asymmetry parameter $|q|$ rapidly decreases in the temperature range up to 100~K and then stays roughly constant up to 300~K. For $T=10$~K it is not obvious that a Fano shape should be preferred compared to a Lorentz shape which indicates that the phonon mode behaves like an independent classical oscillator. For $T\ge50$~K $|q|$ is suppressed and between 50 and 100~K the interaction between the phonon and the electronic continuum becomes relevant. Moreover, as $q$ changes from negative to positive, the energy of the continuum of electronic states exceeds the phonon energy. This result is consistent with the temperature dependence of the Drude weight which dominates the optical conductivity spectra at the energy of the phonon ($\approx24$~meV) at $T=100$ and 300~K only, see Fig. \ref{optcondparc}. Thus, we identify the conduction-electron excitations (the Drude component)  as the origin of the continuum of electronic states. It is worth to note the contrasting lineshape evolution of the data reported by Perucchi et al. \cite{perucchi06a}. In their crystals, due to the presence of a continuum of excitations extending in the spectral range between zero and the energy gap, an asymmetric lineshape with $q\approx-5$ is found for a phonon in the $\mathbf{E}\perp\mathbf{b}$ configuration at low temperatures whereas at high temperatures the phonon shape becomes more symmetric with $q\approx-15$. 

The phonon mode in RuSb$_{2}$ could be fitted well by a Lorentz oscillator at all investigated temperatures. This is expected from the very low charge-carrier density and therefore the interaction between phonons and the conduction-electron continuum should not dominate the phonon relaxation.

\section{Summary}

The optical reflectivity of FeSb$_2$ and RuSb$_2$ is found to be similar only at low temperatures. With increasing temperature up to 300~K the optical properties of both compounds deviate considerably: up to 200~meV the reflectivity of FeSb$_2$ strongly increases due to the growing influence of a Drude component whereas the spectra of RuSb$_2$ show very small changes.   
This disparate behavior indicates a characteristic behavior of two semiconductors with different gap energies. 
We analyzed the semiconducting absorption edge (fundamental absorption) and determined the gap values for FeSb$_2$ (indirect: 31~meV for $T\le100$~K; direct: 130~meV for $T<300$~K) and for RuSb$_2$ (indirect: 260 meV; direct: 790~meV). Moreover, for FeSb$_2$ another indirect gap at 6~meV is found for $T\le50$~K. In this temperature range the charge-carrier density is strongly reduced and the presence of a charge gap of 6~meV seen in resistivity measurements coincides with a colossal Seebeck effect.\cite{bentien07a} Moreover, the decreasing charge-carrier density is consistent with the change in the sign of the Fano asymmetry parameter for the $B_{\rm{1u}}$ phonon mode at 24~meV indicating phonon-electron interactions at high temperatures.

The analysis of the optical spectral weight reveals two energies for FeSb$_2$ at which the spectral-weight loss upon charge-gap formation is recovered: one at 0.25~eV for $T\le100$~K and another one at 1~eV for $T\le300$~K. The fact that these values are much larger than the energies of both the indirect and the direct gaps indicate the relevance of strong electronic correlations for the formation of these gaps. 

A comparison of the experimental optical conductivity with the results from a band-structure calculation supports the presence of strong electron correlations for the formation of (at least) the direct gap in FeSb$_2$.
 
\section*{Acknowledgement}
We acknowledge fruitful discussions with P.~Sun, A.~Bentien and D.~Menzel. This work was a joint studies program of the Institute of Molecular Science (2006) and was partially supported by a Grant-In-Aid for Scientific Research (B) (Grant No. 18340110) from the Ministry of Education, Culture, Sports, Science and Technology of Japan.

\end{document}